\documentclass[]{aa}
\usepackage{graphicx}
\usepackage{txfonts}
\usepackage{natbib}
\bibpunct{(}{)}{;}{a}{}{,}

\begin{document}

\title{A comparison between solar plage and network properties}

\author{D. Buehler \inst{1}$^,$\inst{2}, A. Lagg \inst{2}, M. van Noort \inst{2} \and  S.K. Solanki \inst{2}$^,$\inst{3}}
\authorrunning{D. Buehler, et al.}

\institute{
\inst{1} Institute for Solar Physics, Dept. of Astronomy, Stockholm University, Albanova University Center, SE-10961 Stockholm, Sweden\\
\inst{2} Max Planck Institute for Solar System Research, Justus-von-Liebig-Weg 3, 37077 G\"ottingen, Germany\\
\inst{3} School of Space Research, Kyung Hee University, Yongin, Gyeonggi, 446-701, Korea}

\date{date}

\abstract
{}
{We compare the properties of kG magnetic structures in the solar network and in active region plage at high spatial resolution.}
{Our analysis used six SP scans of the solar disc centre aboard Hinode SOT and inverted the obtained spectra of the photospheric 6302 $\AA$ line pair using the 2D SPINOR code.}
{Photospheric magnetic field concentrations in network and plage areas are on average 1.5 kG strong with inclinations of $10^{\circ}-20^{\circ}$, and have $<400$ m/s internal and $2-3$ km/s external downflows. At the disc centre, the continuum intensity of magnetic field concentrations in the network are on average $10\%$ brighter than the mean quiet Sun, whilst their plage counterparts are $3\%$ darker. A more detailed analysis revealed that all sizes of individual kG patches in the network have 150 G higher field strengths on average, 5$\%$ higher continuum contrasts, and 800 m/s faster surrounding downflows than similarly sized patches in the plage. The speed of the surrounding downflows also correlates with the patch area, and patches containing pores can produce supersonic flows exceeding 11 km/s in individual pixels. Furthermore, the magnetic canopies of kG patches are on average $9^{\circ}$ more horizontal in the plage compared to the network.}
{Most of the differences between the network and plage are due to their different patch size distributions, but the intrinsic differences between similarly sized patches is likely results from the modification of the convection photospheric convection with increasing amounts of magnetic flux.}


\maketitle

\section{Introduction}

The network in the quiet Sun and the plage in solar active regions are readily identified in solar photospheric observations by their excess brightness, particularly in the cores of spectral lines \citep[e.g.][]{jafarzadeh2013}. At the disc centre, the network appears as individual or chains of bright points immersed within dark, downflowing intergranular lanes, and surrounded by bright, upflowing granules \citep{riethmueller2010}. The bright points \citep{berger1996} are additionally affected by a larger scale supergranular convection pattern \citep{leighton1962} that arranges them into a global network. Plage areas typically make up the bulk of the trailing polarity in solar active regions. They also host bright points, but the large magnetic flux in plage areas causes a higher bright point density than in the network as well as the formation of larger structures including pores. If the magnetic flux density is sufficiently large, the bright points merge to fill the intergranular lanes.

Both the network and plage host roughly vertical kG magnetic fields \citep{howard1972,frazier1972,stenflo1973,rabin1992,rueedi1992}, which spatially coincide with the bright points and pores in these areas. They are likely formed by the convective collapse mechanism \citep[e.g.][]{spruit1979}, which concentrates diffuse hG magnetic fields in intergranular lanes into kG features \citep{nagata2008,requerey2014}. Bright points are typically treated as thin flux tubes \citep{spruit1976,defouw1976} where a lateral radiative inflow leads to an excess brightness \citep{deinzer1984,voegler2005}, and pores are regarded as thick flux tubes \citep{knoelker1988,cameron2007} where the lateral radiative heating from the flux tube walls cannot compensate for the lack of convection within the flux tube.

Earlier comparisons indicated that, despite the large differences in magnetic flux between network and plage areas, the properties of the average individual magnetic feature in both areas are nearly the same \citep{stenflo1985,solanki1986}. Only the temperature is on average some 200 K cooler in the typical plage flux tube \citep{solanki1984,solanki1986}, which suggests that the typical flux tube radius is on average larger in the plage than in the network \citep{solanki1992pla}. 

Whilst bright points in the network are routinely observed, studies based on low spatial resolution polarimetric observations often associated them with mere hG magnetic fields \citep{lawrence1993,ortiz2002,kobel2012,yeo2013} or failed to link polarisation signals at the disc centre with continuum intensity enhancements \citep{topka1992}. Spectral analyses, such as the line ratio technique or inversions employing multiple atmospheres per pixel, have yielded kG magnetic fields \citep{stenflo1973,martinez1997,beck2007,viticchie2010,utz2013}, but the individual magnetic features were still spatially unresolved and their true continuum intensity is unknown. Only more recent observations employing $0\farcs1$ resolution \citep{lagg2010,riethmueller2014,kahil2017,kahil2019} or a 2D inversion approach \citep{buehler2015} have obtained kG magnetic fields for bright points  using inversions with only a single atmosphere per pixel. A comparison between G-band bright points intensities by \citet{romano2012} indicated that bright points in the network are systematically brighter than their counterparts in the plage, whereas \citet{ji2016} using lower resolution SDO/HMI observations, concluded the opposite.

Magnetohydrodynamic (MHD) simulations have generally supported the bright thin flux tube picture \citep{stein2006,shelyag2007,yelles2009} and have further suggested that the number of thick flux tubes in plage areas is comparatively small \citep{knoelker1988}. The simulated flux tubes strongly favour a normal orientation with respect to the solar surface \citep{schuessler1986}, and the routinely observed $10^{\circ}-20^{\circ}$ deviation from the normal \citep[e.g.][]{bernasconi1995,martinez1997,jafarzadeh2014,buehler2015} has been interpreted as a sign of buffeting of the flux tubes by turbulent granular convection \citep{steiner1996}. The MHD simulations also reproduced the routinely observed \citep{title1989,title1992,narayan2010,kobel2012,romano2012} 'abnormal' granulation and slight suppression of the granular convection pattern in plage areas compared to the quiet Sun \citep{voegler2005}.

The line-of-sight (LOS) velocity within magnetic elements is on average below $250$ m/s \citep{solanki1986,martinez1997}, but values exceeding 1 km/s have been reported especially when intensity thresholds were used to identify magnetic elements \citep{berger2007,riethmuller2010,romano2012}. Observations of active regions by \citet{langangen2007,cho2010}, and \citet{buehler2015} indicate that LOS velocities within magnetic elements are small, but they are surrounded by fast downflows typically 2 km/s. Whilst it seems logical that magnetic elements in the quiet Sun should also host the fast downflows surrounding them (see observations by  \citet{bonet2008} and \citet{requerey2014} or simulations by \citet{schuessler1984} and \citet{knoelker1988}), no confirmation exists so far. \citet{cho2010} demonstrate that the fast downflows around pores inversely correlate with the pore's continuum intensity, suggesting that the downflows are magnetic field dependent. A similar relation for magnetic elements in the network or plage is still outstanding. 

Older comparisons between magnetic elements in the network or plage often focussed on individual features such as isolated bright points or suffered from the low spatial resolution of the data \citep[e.g.][]{solanki1984,solanki1985,stenflo1985,solanki1986}. Other studies compared the properties of these magnetic elements more generally without distinguishing between feature sizes ranging from isolated bright points to strings of bright points to minor pores. In this investigation we attempt to provide a more comprehensive comparison in terms of their various properties such as magnetic field strengths, continuum intensities, or LOS velocities as well as their dependence on feature size. We employ stable, high resolution Hinode observations inverted using the 2D inversion approach introduced by \citet{vannoort2012} that allows us to fit each pixel using a single atmosphere and, hence, to fit the Stokes profiles in the individual magnetic elements in both regions. Our results may aid future investigations such as irradiance studies \citep[e.g.][]{foukal1988,krivova2003} and research into chromospheric phenomena \citep[e.g.][]{tsiropoula2012}.

\section{Observations \& analysis}
This study uses six data sets recorded by the spectropolarimeter (SP), which is part of the Solar Optical Telescope (SOT) \citep{tsuneta2008sot,suematsu2008sot,ichimoto2008sot,shimizu2008sot} aboard the Hinode satellite \citep{kosugi2007}. The SP records the photospheric and magnetically sensitive 6301 $\AA$ and 6302 $\AA$ Fe I line pair. The majority of the data sets were recorded in the 'normal' mode, which prescribes a total exposure time of 4.8 s per slit position at a spatial sampling of 0\farcs16. At each slit position the four Stokes parameters $I$, $Q$, $U$, and $V$ were recorded with a noise level of $1.1 \times 10^{-3}$ $I_c$. Two data sets have exposure times of 9.6 s and 12.8 s per slit position, and thus noise levels of $8 \times 10^{-4}$ $I_c$ and $7 \times 10^{-4}$ $I_c$, respectively. Each data set was subsequently reduced using $sp\_prep$ \citep{lites2013}. 

The six data sets analysed here are listed in Table \ref{SPtable} and shown in Figs. \ref{overviewNTPcont} and \ref{overviewNTPmag}. All depict regions close to the disc centre at the time of observation. The two plage data sets both contain sunspots, which were cut out before the analysis to avoid contamination of the results by the magnetic fields from the sunspot canopies, hence, all pixels showing extended large scale homogeneous fields at $\log(\tau)=-2$ were removed. The inversion results of the first plage data set in Table \ref{SPtable} were also analysed in \citet{buehler2015,buehler2016}. Data set 3, which covers an extended area of quiet Sun containing numerous network elements, was cut into three separate strips that were inverted independently. Multiple datasets were employed to improve the statistical analysis in particular for quiet Sun areas.

Both spectral lines in all data sets were inverted using the SPINOR code \citep{frutiger2000} in the 2D spatially coupled mode \citep{vannoort2012} and kept the spatial sampling of 0\farcs16. The inversion compensates for the degradation caused by the telescope diffraction during the inversion and thereby eliminates the need to introduce a second atmosphere in each pixel that functions as a straylight component. Consequently the Stokes spectra of each pixel are described in terms of a single atmospheric component. The amplitude and area asymmetries in the spectra, indicative of LOS gradients in a pixel's atmosphere \citep[e.g.][]{solanki1993}, were accounted for in the inversion by setting three nodes in optical depth at $\log(\tau)=0$, $-0.8$ and $-2$. The atmospheric parameters, the temperature, $T$, the magnetic field strength, $B$, the inclination of the field relative to the LOS, $\gamma$, the field azimuth in the LOS coordinate system, $\psi$, the LOS velocity, $v$, and microturbulence, $\xi$, could be altered at these nodes. After solving the radiative transfer equation with the STOPRO routines \citep{solanki1987phd}, which are part of the SPINOR code, the synthesised spectra of a pixel's model atmosphere could be compared to the corresponding observed spectra. A Levenberg-Marquardt minimisation was employed to iteratively modify a pixel's atmospheric parameters until a close match between the observed and synthesised spectra was obtained.    

Following the inversion, the LOS magnetic field inclinations and azimuths were converted to local solar coordinates and the $180^{\circ}$ ambiguity was resolved using the method outlined in \citet{buehler2015}. The resolved magnetic field inclinations are labelled $\Gamma$.

The LOS velocities were calibrated by assuming that the velocities within pores are on average zero, which led to a 200 m/s correction. In data sets that lacked pores, an average, convective blue-shift corrected quiet Sun profile was used instead, which led to a similar correction.  

\begin{table}
\caption{Analysed SP scans.}              
\label{SPtable}      
\centering                                      
\begin{tabular}{c c c c c}          
\hline\hline                        
Date & Type & $\mu$ & Exposure [s] & Area$^{a}$ $['']$\\    
\hline                                   
     2007 04 30 & Plage & 0.94 & 4.8 & 48 $\times$ 160\\
     2007 05 18 & Plage & 0.99 & 4.8 & 80 $\times$ 96\\
     2007 02 18 & Network & 0.99 & 4.8 & $144^{\rm{b}}$ $\times$ 160\\      
     2007 09 10 & Network & 0.99 & 4.8 & 60 $\times$ 128\\
     2007 09 10 & Network & 0.99 & 9.6 & 60 $\times$ 128\\
     2007 09 10 & Network & 0.99 & 12.8 & 60 $\times$ 128\\
\hline                                             
\end{tabular}
\tablefoot{a: Denotes inverted area. b: Composed of three separately inverted strips.}
\end{table}    

\section{Results}
\begin{figure}
       \centering
        \includegraphics[width=8cm]{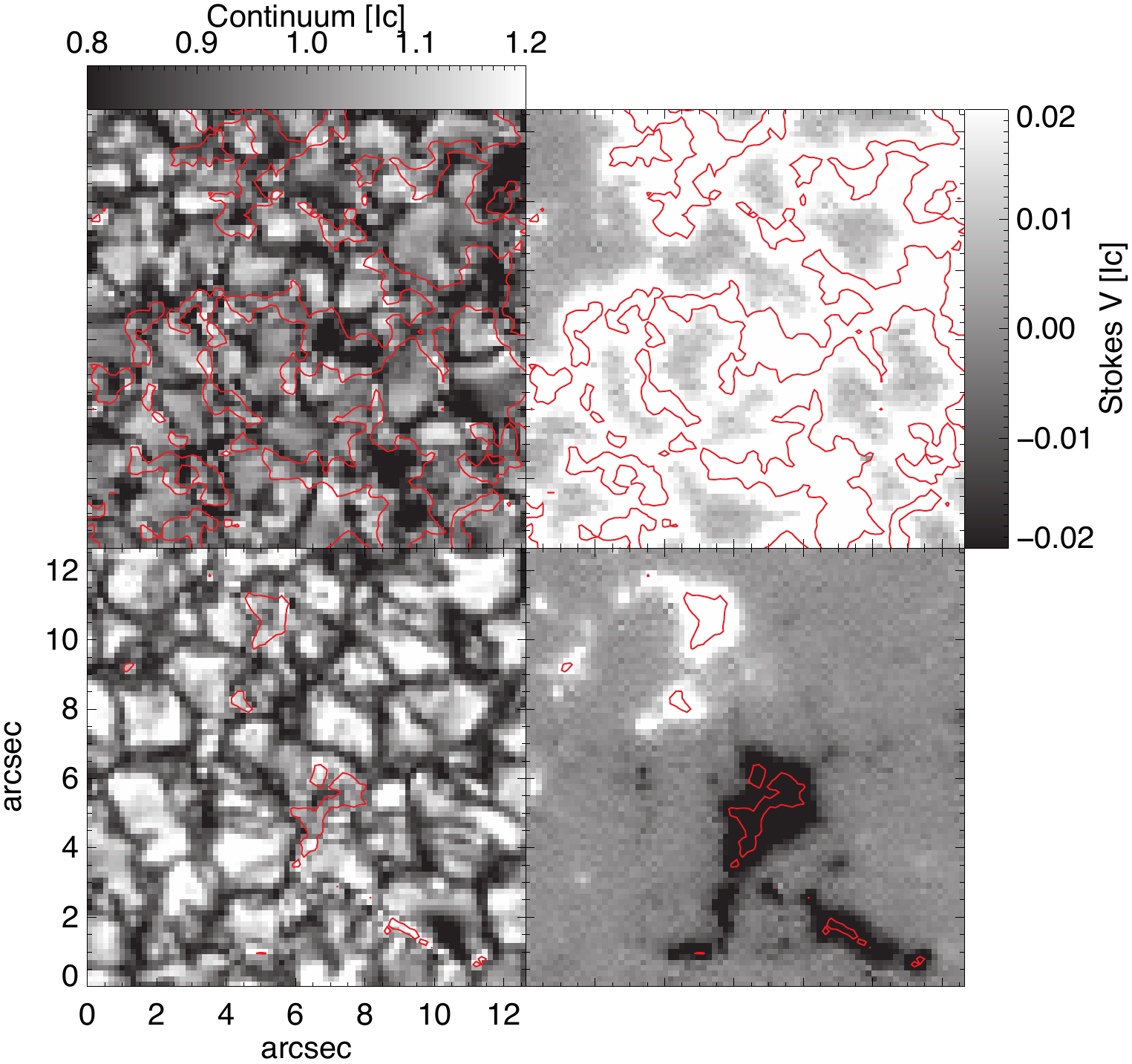}
        \caption{Continuum intensity and Stokes $V$ images of a plage region, top row, and a network region, bottom row. The red contour lines encompass kG magnetic fields at $\log(\tau)=-0.8$. The Stokes $V$ images display the signal amplitude at $6301-0.06$ $\AA$ and have been saturated at a level of $2\%$.}
         \label{Overview}
         \end{figure}

Stokes $I$ continuum and $V$ maps of a typical observation of a plage and network region are provided in Figure \ref{Overview}. Numerous bright points can be seen within intergranular lanes in both continuum images and small pores can additionally be spotted in the plage image. The Stokes $V$ images display the strong polarisation signals produced by these structures and indicate the presence of magnetic fields within them. The Stokes $V$ patches suggest that the area covered by magnetic fields far exceeds the size of the associated bright points/pores. However, large parts of these Stokes $V$ patches do not contain kG magnetic fields in the lower photosphere as the red contour lines demonstrate, which enclose kG fields at $\log(\tau)=-0.8$. The patches retrieved from the contour lines, where the field penetrates the $\tau=1$ layer, closely follow the outline of bright points and pores in the continuum images.

\begin{figure}
       \centering
        \includegraphics[width=8cm]{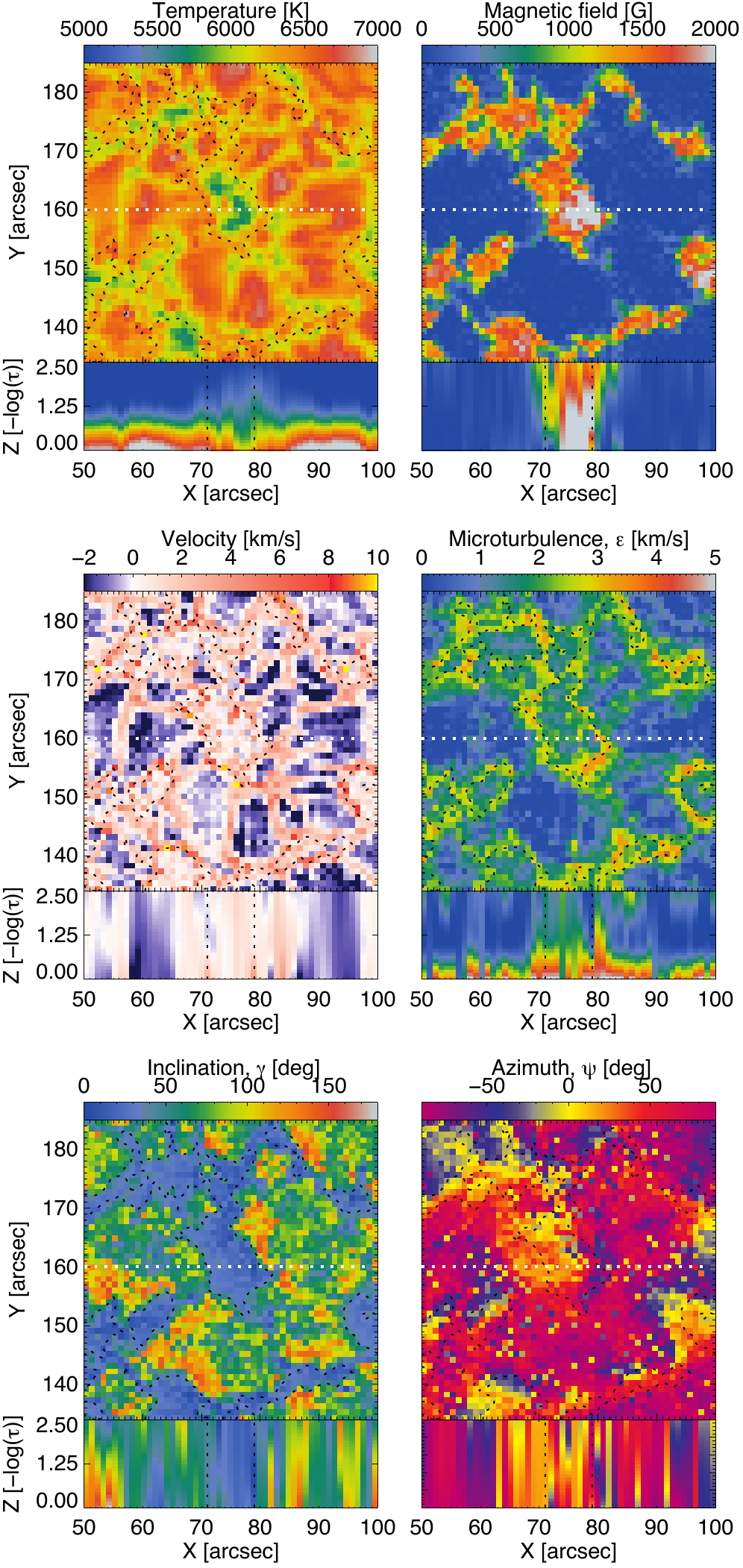}
        \caption{Inversion results of a typical plage region. The temperature and LOS velocity images depict the $\log(\tau)=0$ layer whereas all other panels display the $\log(\tau)=-0.8$ layer. The black dotted contour lines indicate kG magnetic fields at $\log(\tau)=-0.8$ and the white lines display the location of the vertical cuts. The mean field strength of the image at $\log(\tau)=-0.8$ is 450 G.}
         \label{Pslice}
         \end{figure}

\begin{figure}
        \centering
        \includegraphics[width=8cm]{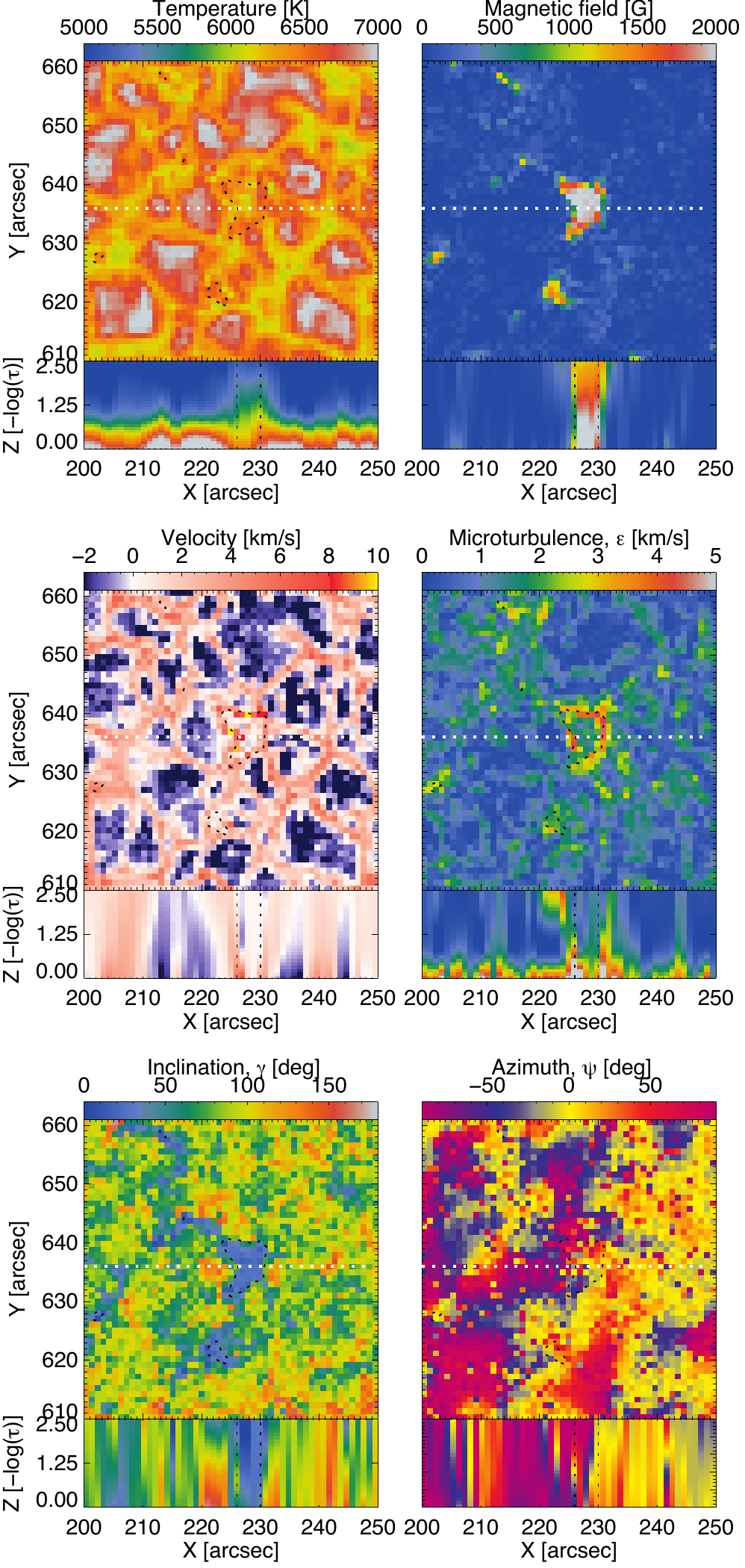}
        \caption{Inversion results of typical network region similar to Figure \ref{Pslice}. The mean field strength of the image at $\log(\tau)=-0.8$ is 110 G.}
         \label{NTslice}
         \end{figure}

The pixels located outside a contour line surrounding a Stokes $V$ patch generally belong to the magnetic canopy as demonstrated by the magnetic field maps and slices in Figs. \ref{Pslice} and \ref{NTslice}. These canopy magnetic fields expand over the nearly field free convection cells underneath and do not modify the continuum intensity at the disc centre. They are largely responsible for the size discrepancy displayed by magnetic features between the continuum and Stokes $V$ images. 

Table \ref{comp} lists the average atmospheric values for a typical pixel in a plage and network magnetic flux concentration at the three $\log(\tau)$ nodes. Pixels containing canopy fields, that have $B < 1$ kG at $\log(\tau)=-0.8$, and sunspots found in the plage data sets were excluded. Pores, which are frequently found in plage areas (e.g. see Figs. \ref{Overview} and \ref{Pslice}), are included in these averages.
 
Both the average network and plage pixels display a characteristic drop in temperature with height whilst the network features are consistently hotter by at least 200 K compared to the plage. The average temperatures listed for $\log(\tau)=-2$ would be somewhat cooler if the canopy pixels were included, which are typically 100 K cooler than the values listed in Table \ref{comp}. The plage and network temperature stratifications match the semi-empirical PLA atmosphere \citep{solanki1992}, modelling a zeroth order thin flux-tube fairly well within the constraints imposed by a three $\log(\tau)$ node atmosphere. Furthermore, the temperature stratification in the plage is steeper the smaller the magnetic feature, and the smallest features match the PLA model more closely than the largest features, which contain pores. The temperature stratification of the PLA model also fits the temperature stratification of flux tubes produced by 3D MHD simulations \citep{vitas2009} reasonably well. Since the output of the MuRAM 3D MHD simulations has proved very successful in reproducing a large range of observations, including the variation of total solar irradiance \citep{yeo2017}, this suggests that a comparison with the PLA model is reasonable. The network temperature stratification by comparison displays no patch area dependence as it does not contain pores.

The average magnetic field strengths in Table \ref{comp} are nearly identical for the network and plage.  They decrease with height due to horizontal pressure balance and the magnetic features expand to conserve magnetic flux. The expansion of isolated magnetic features in the network and plage both approximately follow the zeroth order thin flux-tube PLA and NET models \citep{solanki1986}. We selected in total twenty isolated magnetic features from the network and plage data sets and measured their expansion according to the method described in \citet{buehler2015}. The hotter flux tubes in the network should, according to theory, expand more rapidly than their cooler plage counterparts. Unfortunately, the scatter within the set of selected plage and network features prevented us from discerning this difference. 

The network and plage magnetic features display nearly vertical magnetic fields throughout the photosphere with a mean inclination of $20^{\circ}$. The mode inclinations are within the $10^{\circ}-20^{\circ}$ range for both the network and plage. 

The LOS velocities in the network and plage are on average nearly at rest with downflows of no more than 400 m/s across nearly all $\log(\tau)$ layers at which the measurements can be considered reliable. The network appears to have slightly faster downflows than the plage, particularly in the $\log(\tau)=0$ layer. However, the fast downflows in this layer are contaminated by the even faster convective downflows surrounding the magnetic features. The kG contour line at $\log(\tau)=-0.8$ employed to select magnetic features fails to fully exclude the fast downflows at $\log(\tau)=0$ around the magnetic features due to the magnetic field expansion with height. This effect is less impacting in the plage due to the presence of larger features.

Considerable microturbulent velocities are required to successfully fit the plage and network spectra. The microturbulent velocities decrease with height within the magnetic features and are always larger than in the average quiet Sun except for the $\log(\tau)=0$ layer. The network displays somewhat larger microturbulent velocities than the plage across all $\log(\tau)$ layers. 

\begin{table}
\caption{List of average plage, network, and quiet Sun values}              
\label{comp}      
\centering                                      
\begin{tabular}{l c c c | c c}          
\hline\hline                        
Parameters & $-\log(\tau)$ & Plage & Network & QS & QS\\
&&&& G & I \\  
\hline\hline                                   
    $T$ [K]                & 2.0    & 5000 & 5390 & 4910 & 4930\\      
                                & 0.8    & 5610 & 5960 & 5460 & 5300\\
                                & 0.0    & 6200 & 6460 & 6730 & 6310\\
\hline                      
    $B$ [G]                & 2.0    & 1200 & 1220 & 50 & 50\\
                                & 0.8    & 1530 & 1520 & 50 & 60\\
                                & 0.0    & 1630 & 1680 & 80 & 150\\
\hline                      
 $\Gamma$ [$\circ$] & 2.0    & 22 & 17 & 90 & 90\\
                                & 0.8     & 20  & 21 & 90 & 90\\
                                & 0.0     & 25  & 32 & 90 & 90\\
\hline   
   $v$ [km/s]              & 2.0    & -0.05 & 0.22 & -0.25 & 0.48\\
                                 & 0.8    & 0.06 & 0.39 & -0.62 & 0.78\\
                                 & 0.0    & 0.37 & 1.18 & -0.77 & 1.90\\
\hline
   $\xi$ [km/s]            & 2.0    & 1.21 & 1.54 & 0.43 & 0.23\\
                                 & 0.8    & 1.89 & 2.10 & 1.05 & 0.67\\
                                 & 0.0    & 3.26 & 3.72 & 3.32 & 3.07\\
\hline\hline
                                             
\end{tabular}
\tablefoot{The plage and network values represent spatial average over all kG cores of the magnetic structures. See contour lines in Figs. \ref{Overview}, \ref{Pslice} and \ref{NTslice} for reference. The quiet Sun averages are divided between granular (G) and intergranular (I) areas found in internetwork areas.}
\end{table} 

Table \ref{SPtable} indicates that we employed two data sets with exposure times longer than $4.8$ s. Given that magnetic field concentrations in the network are on average smaller and closer to the spatial resolution limit than in the plage (see Figure \ref{Overview}), data sets with lower noise levels benefit the analysis of the network in particular. However, an increased exposure time raises the risk of temporally smearing features and therefore altering their apparent properties. This could be due to proper motions of the features, or, for instance, due to integrating over waves within the features. We decided to test for this possibility by selecting sub-FOVs ($70\times70$ pixels) in each data set and imposed that each sub-FOV needs to have the same flux density, in this case $1400$ G at $\log(\tau)=-0.8$ averaged over the pixels hosting kG fields. The sub-FOV we selected mainly contained small magnetic structures to make them as similar to each other as possible and in the following we compare some of the results obtained for the different exposure times. The mean microturbulence for kG pixels in each sub-FOV at $\log(\tau)=-0.8$ is $2.21\pm0.05$ km/s, $2.25\pm0.05$ km/s, and $2.27\pm0.06$ km/s for the exposure times $4.8$ s, $9.6$ s, and $12.8$ s, respectively, where the error is the error of the mean. The mean microturbulence at $\log(\tau)=-0.8$ in a similar sub-FOV in the plage (data sets 1 and 2 in Table \ref{SPtable}) is $2.26\pm0.05$ km/s and $2.21\pm0.07$. The temperatures at $\log(\tau)=0$ by comparison are $6530\pm70$ K, $6580\pm60$ K, and $6590\pm60$ K for the network data sets with exposure times of $4.8$ s, $9.6$ s, and $12.8$ s, respectively, and $6390\pm60$ K and $6380\pm40$ K for the two plage data sets. Therefore, it becomes apparent that the different exposure times have a negligible effect on the average values and, furthermore, the intrinsic differences between network and plage are independent of the selected data set and the inversion routinely converges on statistically identical values. Other $\log(\tau)$ heights and parameters display a qualitatively similar behaviour. The averages calculated here are different from the ones in Table \ref{comp} due to the small FOV which is too small to sample the average quiet Sun, as well as by our decision to select regions hosting only small magnetic features for this particular comparison.

In the following, we not only examine individual pixels but also entire magnetic patches. The patches are similar to the features outlined by the contour lines in Figs. \ref{Overview}, \ref{Pslice}, and \ref{NTslice}, that is the number of connected pixels that contain kG magnetic fields at $\log(\tau)=-0.8$. From Figure \ref{Overview} it is clear that many patches contain several bright points or even small pores. The multi-cored nature of these patches can also be seen in the inversion results displayed in Figs. \ref{Pslice} and \ref{NTslice}.      

\subsection{Continuum intensity and magnetic field}
\begin{figure}
        \centering
        \includegraphics[width=8cm]{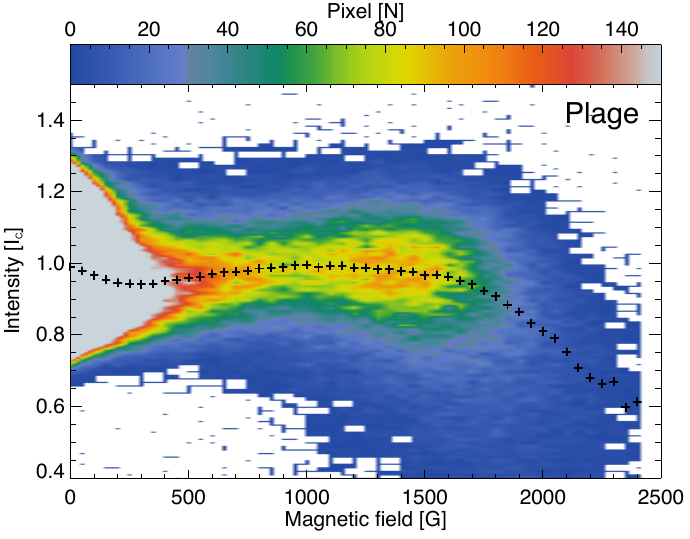} 
        \includegraphics[width=8cm]{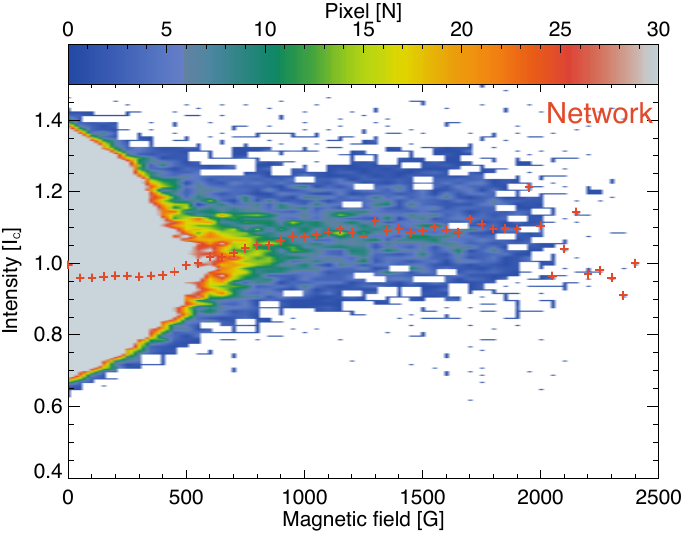}
        \caption{2D histogram of magnetic fields at $\log(\tau)=-0.8$ and continuum intensity in the plage and network. The black and red plus symbols indicate the mean continuum intensity at 50 G intervals in the plage and network, respectively.}
         \label{BT}
         \end{figure}

Figure \ref{BT} displays the relationship between the magnetic field strength at $\log(\tau)=-0.8$ and the normalised continuum intensity for all pixels in the plage and network images excluding sunspots. The continuum intensity was determined separately for each data set using an area that approximates the quiet Sun (i.e. with $B<1$ kG at $\log(\tau)=-0.8$). At each 50 G interval the continuum intensity distribution was fitted with a Gaussian. The plus symbols indicate the average continuum intensities from those fits. Magnetic fields below the equipartition field strength ($\sim$ 450 G) reside in both hot granules and cool intergranular lanes. Pixels with higher hG fields strengths progressively become brighter. In plage areas, pixels with kG field strengths achieve on average a continuum intensity similar to the quiet Sun before rapidly reducing in brightness due to the numerous pores present in the plage, which contain the strongest magnetic fields. The network is characterised by the general absence of pores, which allows even pixels approaching field strengths of 2 kG to be on average 10\% brighter than the mean quiet Sun.
         
\begin{figure}
        \centering
        \includegraphics[width=8cm]{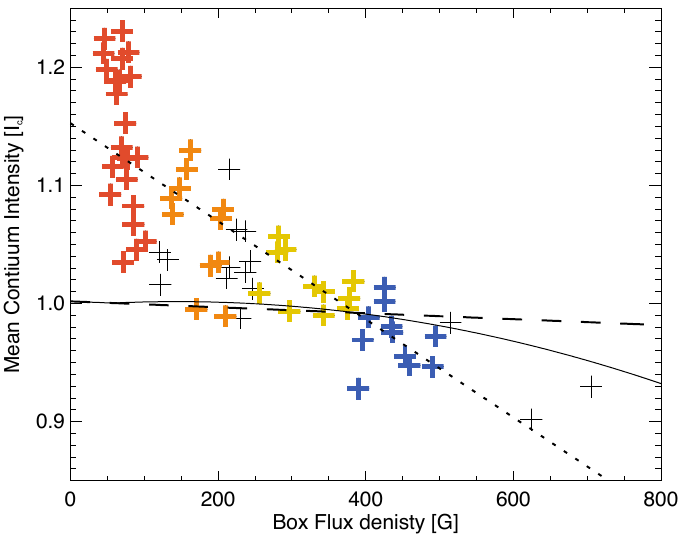}
        \caption{Correlation between an area's flux density and its mean continuum intensity. The plus symbols show the mean continuum intensity of kG pixels within each box. The coloured crosses indicate the boxes that were summed for the similarly coloured distributions in Figs. \ref{MagI}, \ref{generalB}, and \ref{CanIncl}. The dotted line is a linear fit to the mean continuum intensity of kG pixels. The dashed line displays a linear fit to the mean continuum intensity of sub-kG pixels. The solid line shows the mean continuum intensity of all pixels in a box.}
        \label{generalI}
        \end{figure}

Despite the clear differences between the network and plage displayed in Figure \ref{BT}, the transition from network to plage appears to be smooth and continuous. We mimicked this transition by indiscriminately dividing the SP scans listed in Table \ref{SPtable} into smaller boxes, which were then sorted by the mean flux density of each box at $\log(\tau)=-0.8$. The box size used in the following was 50 by 50 pixels or $8'' \times 8''$ and the results are given by the plus signs in Figure \ref{generalI}. The impact of larger or smaller box sizes were tested, but did not qualitatively alter the results.  

The rapid darkening of kG fields with increasing box flux density is displayed by the dotted line in Figure \ref{generalI}, which is a linear fit to the symbols plotted in the figure and takes the form $I_c =  1.14\pm0.01 - 4.0\pm0.4\times10^{-4}B$, where $B$ is the field strength in Gauss. It indicates that above a magnetic flux density of $\sim400$ G in the $8'' \times 8''$ box the kG features are on average dark. The mean continuum intensity of pixels harbouring sub-kG fields at $\log(\tau)=-0.8$, shown by the dashed line from the equation $I_c = 0.998\pm0.002 -1.5\pm0.5\times10^{-5}B$, also steadily decreases with increasing flux density in the boxes albeit much more gradually. In the strongest plage regions with box flux densities of 700 G the granular convection cells have a mean continuum intensity $1\%$ below the quiet Sun level.  The solid line represents the mean intensity of the whole box weighted by the relative contributions of the dotted and dashed lines. The solid line has a maximum near 150 G and 200 G with a mean intensity of 0.1\% above the the quiet Sun at which point 5\% of pixels in the box contain kG fields. At a box flux density of 600 G nearly 25\% of the box is filled with kG fields.

        \begin{figure}
        \centering
        \includegraphics[width=8cm]{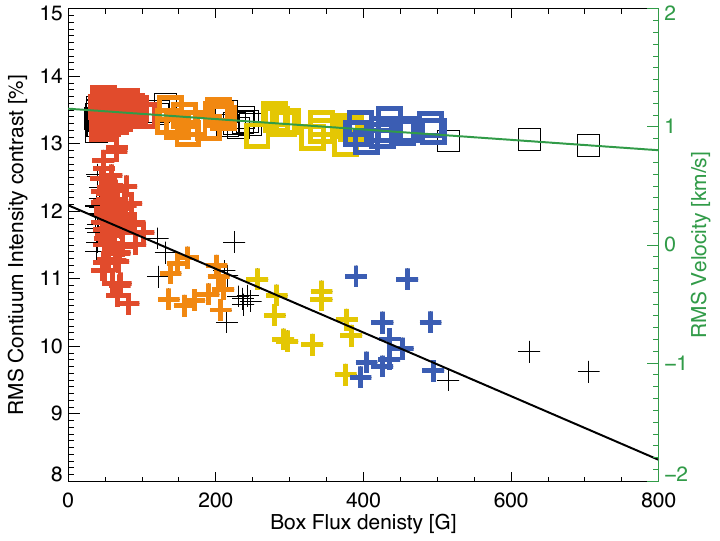}
        \caption{Plus symbols: Scatterplot between an area's flux density and rms continuum intensity contrast in sub-kG pixels. The black solid line is a linear fit to the data. Square symbols: Scatterplot between an area's flux density and rms LOS velocity in sub-kG pixels. The green solid line is a linear fit to the data. The colour scheme is identical to Figure \ref{generalI}}
        \label{contrast}
        \end{figure}

The rms continuum intensity contrast of the sub-kG pixels also decreases with increasing flux density. Figure \ref{contrast} illustrates that the inversions achieve a contrast of $11.8\%$ in the quiet Sun, which is nearly $3\%$ below the contrast of $14.4\%$ derived from MHD simulations at the same wavelength \citep{danilovic2008}. It indicates that residual scattered light may be present in the data and that the finite spatial resolution of SP cannot resolve all the fine structure in the QS, including the smallest convective structures and the smallest magnetic features (see \citep{riethmueller2014}). The 3D MHD simulations by \citet{criscuoli2013} and \citet{criscuoli2014} and in particular the simulations by \citet{roehrbein2011} and \citet{danilovic2013} point in a similar direction in that our inversions can retrieve the general dependence between magnetic field strength and continuum intensity. However, in particular the smallest kG features tend to have field strengths (see Figure \ref{BT}) and continuum intensity contrasts (see Figure \ref{generalI}) that are below those obtained from the simulations. This may indicate that these features are not completely resolved in the observations. 

The rms contrast in the granulation drops below $10\%$ in regions containing the strongest plage and the reduction in the contrast is mainly driven by the gradual disappearance of bright, hot granules, whilst the intergranular lanes do not show changes in their continuum intensity with increasing flux density. In addition, the granulation in areas with flux densities of 400 G has rms LOS velocities that are 200 m/s lower than in the quiet Sun, which is also shown in Figure \ref{contrast}.

        \begin{figure}
        \centering
        \includegraphics[width=8cm]{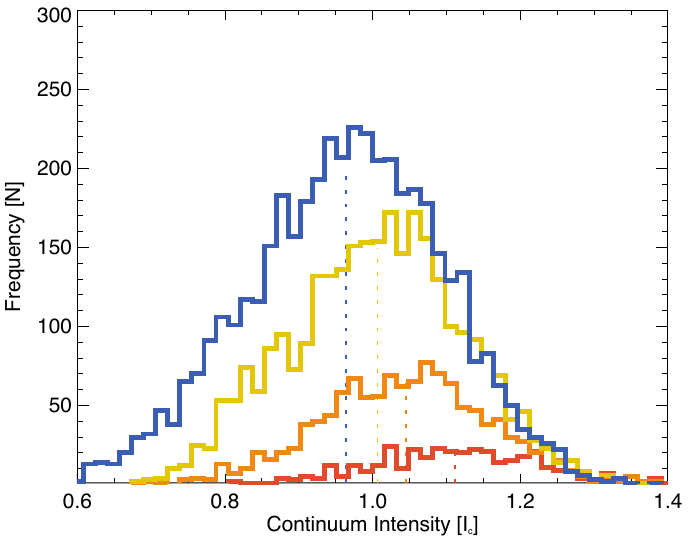}
        \caption{Histograms of continuum intensity of kG pixels. The colour scheme refers to areas of different flux density and is identical to Figure \ref{generalI}. The dotted lines indicate the mean continuum intnesity of each distribution.}
        \label{MagI}
        \end{figure}

Pixels hosting kG fields at $\log(\tau)=-0.8$ display a marked change in their continuum intensities with increasing field strength according to Figure \ref{BT}. We can further illustrate this phenomenon by combining boxes of similar flux density to produce continuum intensity histograms, which are depicted in Figure \ref{MagI}. The coloured symbols in Figs. \ref{generalI} indicate which boxes were combined to create each histogram in Figure \ref{MagI}. The red histogram in Figure \ref{MagI} represents the quietest Sun in our data sets and consequently the majority of kG pixels is brighter than the mean quiet Sun. The blue histogram in the same figure represents the strongest plage boxes and is the only distribution that includes pores. These pores produce a tail of dark pixels in the continuum intensity distribution. The darkest pore pixel has a continuum intensity of only 0.28 $I_c$, but the majority of pores, which are typically embedded within larger magnetic patches, have continuum intensities between 0.5 and 0.7 $I_c$. The mean continuum intensity of the red histogram is $12\%$ brighter than the mean quiet Sun, whilst the blue histogram has a mean continuum intensity that is $3\%$ darker than the mean quiet Sun. However, as the pore-less orange and yellow distributions in Figure \ref{MagI} demonstrate, the progressive decrease in the mean continuum intensity of these distributions is not primarily caused by the addition of pores.

        \begin{figure}
        \centering
        \includegraphics[width=8cm]{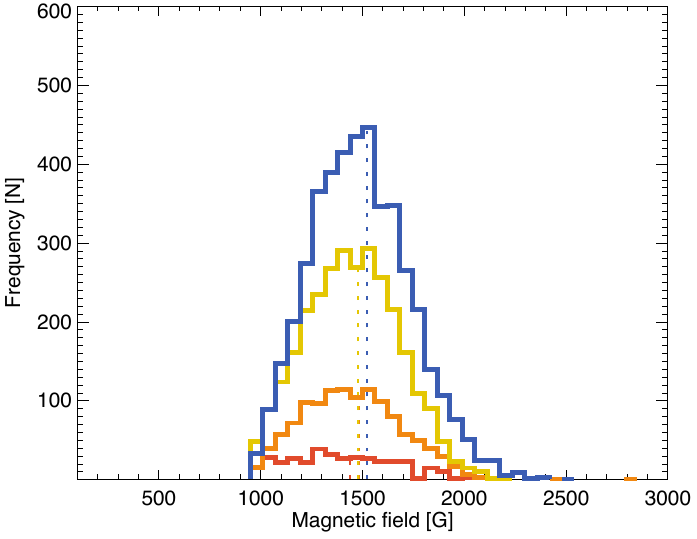}
        \caption{Histograms of magnetic field strength of kG pixels at $\log(\tau)=-0.8$. The colour scheme refers to areas of different flux density and is identical to Figure \ref{generalI}. The dotted lines indicate the mean field strength of each distribution.}
        \label{generalB}
        \end{figure}
        
Whilst the mean continuum intensity of kG fields in an area gradually changes with increasing flux density as seen in Figure \ref{MagI}, the mean magnetic field strength of the kG fields appears not to change, as displayed in Figure \ref{generalB}. All the magnetic field distributions have a mean field strength of $1500\pm50$ G except for the red one, which has a mean field strength of 1350 G. The lower average field strength of this distribution likely stems from the low flux density boxes sampled from the internetwork, which are too quiet to give an accurate representation of the quiet Sun network, which does host an average field strength of 1500 G in kG pixels according to Table \ref{comp}. Apart from increasing the number of kG fields, higher flux densities appear to only gradually widen the distributions shown in Figure \ref{generalB}.  
         
        \begin{figure}
        \centering
        \includegraphics[width=8cm]{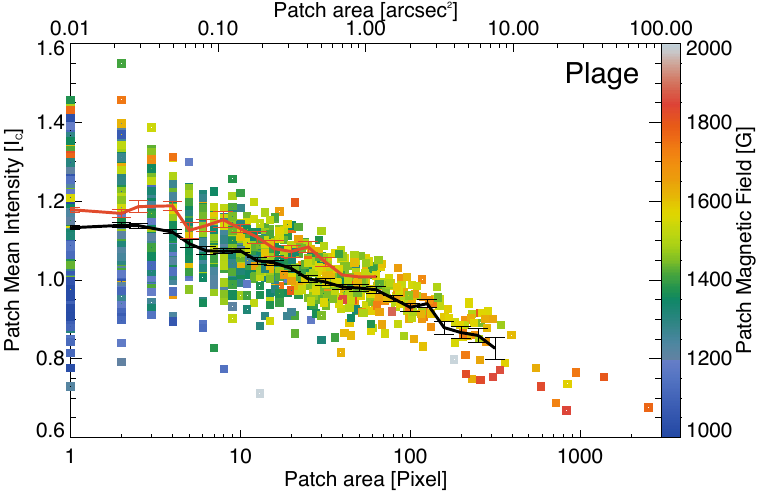}
        \includegraphics[width=8cm]{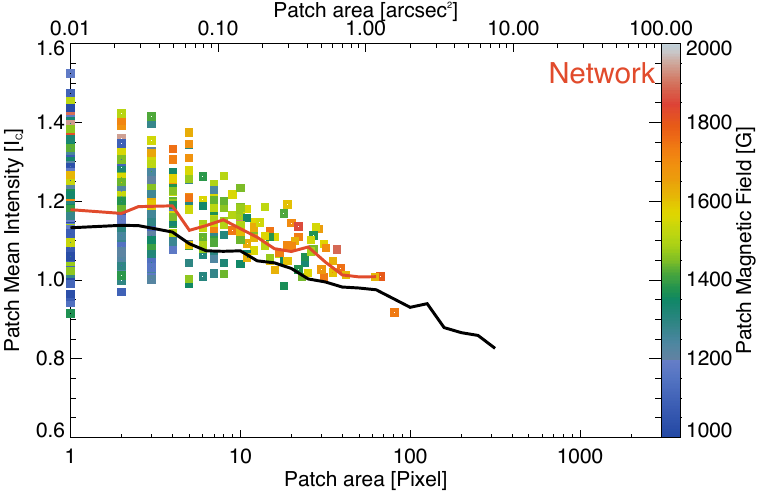} 
        \caption{Scatterplot of patch areas of kG features and their mean continuum intensity in the plage, top, and the network, bottom. The black and red solid lines indicate the plage and network mean continuum intensities respectively. The mean intensities were calculated using ten logarithmic bins per decade of patch area. The error bars refer to the error of the mean.}
        \label{BTpatch}
        \end{figure}
        
The distributions displayed in Figure \ref{BT} can be further refined by grouping pixels with $B$ > 1 kG at $\log(\tau)=-0.8$ into patches. All kG pixels linked by at least a single adjacent kG pixel (i.e. within a one pixel radius) were combined into a single patch. Some example patches are outlined by the contours in Figure \ref{Overview}. The procedure produced 592 network patches and 1994 plage patches in total. The resulting scatterplots of mean patch intensity vs. patch area are displayed in Figure \ref{BTpatch}. The figure paints a familiar picture of small bright magnetic patches and larger darker ones. Patches, larger than 1000 pixels, are composed of several pores connected by clusters of bright points, one of which is displayed in Figure \ref{Pslice}. As a reference, round feature with an equivalent area of $1'' \times 1''$ would possess a radius of 410 km, but the vast majority of the selected patches are multi cored and irregular, allowing even relatively large patches to possess a continuum intensity in excess of the mean quiet Sun. The solid lines in Figure \ref{BTpatch} indicate the mean intensities of the two distributions and reveal that plage patches are on average 5$\%$ darker than their network counterparts regardless of patch size.    
  
        \begin{figure}
        \centering
        \includegraphics[width=8cm]{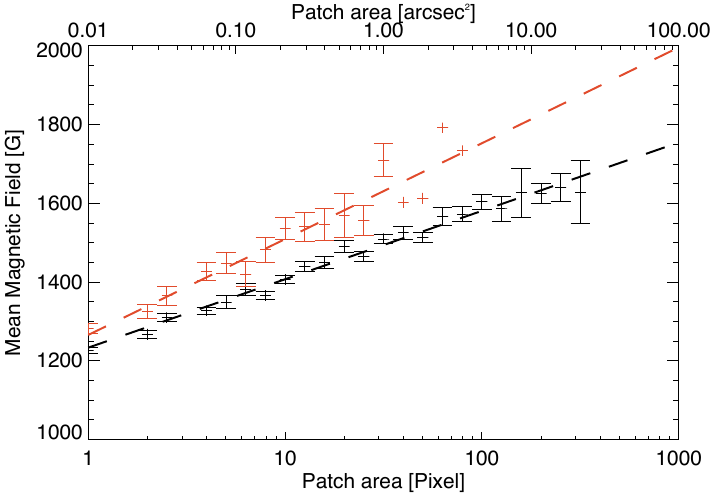}
        \caption{Scatterplot of patch areas of kG features and their mean magnetic field strength at $\log(\tau)=-0.8$. The black symbols refer to patches in the plage and the red symbols are from the network. The error bars refer to the error of the mean. The mean magnetic field strengths were calculated using ten logarithmic bins per decade of patch area.}
        \label{Field}
        \end{figure}          
        
The average magnetic field strength of a patch increases with size as suggested by the colour-coded symbols in Figure \ref{BTpatch}. We proceeded by dividing each decade of patch areas into ten bins and calculated an average patch area field strength for each bin. Figure \ref{Field} displays the resultant relationship between the sizes of patches and their average field strengths for both network and plage patches. The magnetic fields in the network are on average $\sim$ 100 G stronger than in the plage for any given patch area. However, the network lacks the large patches that host the highest mean magnetic field strengths commonly found in the plage. Both effects appear to compensate each other given that the mean field strength all kG pixels found in both network and plage areas is 1.5 kG at $\log(\tau)=-0.8$ (see Table \ref{comp}). The difference between plage and network field strengths persists for all $\log(\tau)$ layers and also when the $\langle B_{max} \rangle$ at a given patch area is used instead of $\langle B \rangle$ as displayed in Figure \ref{Field}.

\begin{figure}
        \centering
        \includegraphics[width=8cm]{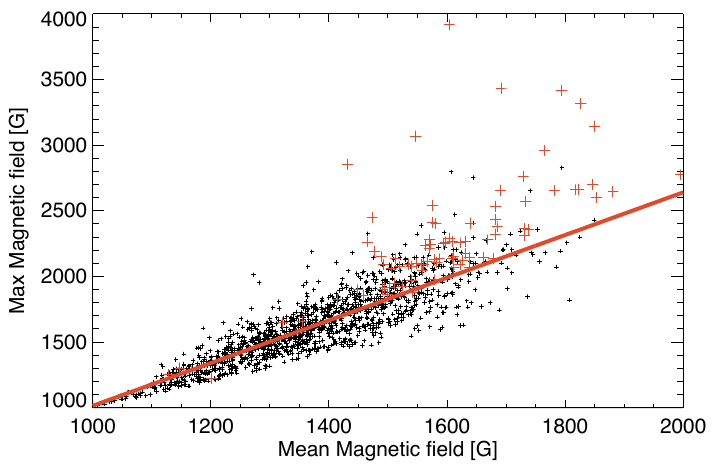}
        \caption{Correlation between mean and maximum magnetic field strength of kG features in plage areas at $\log(\tau)=-0.8$. The black dots belong to patches without pores and the solid red line represents a linear fit. The red crosses represent patches that contain pore pixels.}
        \label{Bmax}
        \end{figure}

Figure \ref{Bmax} indicates that the maximum field strength in a patch is linearly dependent on the mean field strength of a patch. The linear relationship between the mean and maximum field strength of a patch begins to break down in patches containing pores. The maximum field strength in these patches can reach as high as 3.5 kG depending on the size of the pore within the magnetic patch and they generally have a mean fields strength $>1500$ G. One pore attains a maximum field strength of nearly 4 kG in a single pixel at $\log(\tau)=-0.8$, which is situated at the edge of the pore and features fast co-spatial downflows exceeding 10 km/s. The pore is likely part of an emerging flux region as kG patches of opposite polarity were found nearby. Magnetic loops connecting the opposite polarities are present in the  magnetic field and inclination maps obtained form the inversion. The 4 kG pixel is found at one of the foot points of the magnetic loops. 
        
\begin{figure}
        \centering
        \includegraphics[width=8cm]{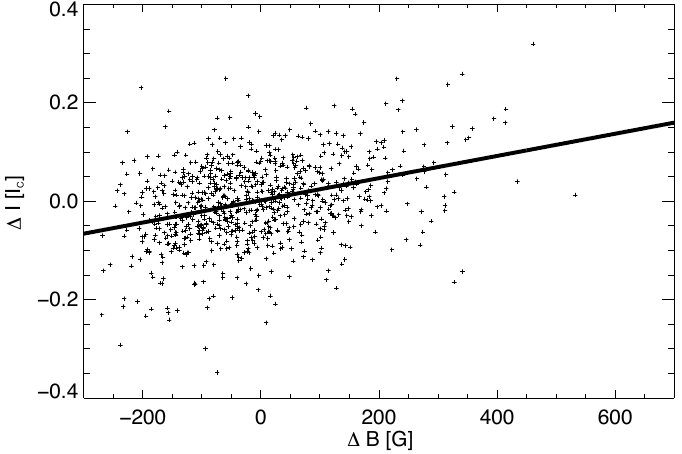}
        \caption{Correlation between continuum intensity and magnetic field strength of kG features < 20 pixels in size after normalising for their mean continuum intensity, shown in Figure \ref{BTpatch} and their mean magnetic field strength in Figure \ref{Field}. The correlation coefficient is 0.4 and the solid line represents a linear fit to the data.}
        \label{deltaI}
        \end{figure}

Small patches display a large variation of continuum intensities ranging from 0.8 - 1.5 $I_c$ according to Figure \ref{BTpatch}. A closer inspection of the average magnetic field strengths of these patches reveals that brighter patches also tend to have a higher magnetic field strength (see colour coding in Figure \ref{BTpatch}). We isolated this effect by normalising patches of a given patch area by their average mean intensity (see black line in Figure \ref{BTpatch}) and their average magnetic field strength (see black line in Figure \ref{Field}). The result is displayed in Figure \ref{deltaI} and shows a weak correlation ($r=0.4$) between continuum intensity and magnetic field strength. The relation displayed in Figure \ref{deltaI} is identical for the network and plage and holds for patch areas of up to $\sim$20 pixels. The slope of the regression becomes progressively flatter for larger patches and is absent in patches containing pores. Since the brightness of pores correlates with feature size and maximum field strength according to Figure \ref{BTpatch}, the breakdown of the relation in Figure \ref{deltaI}, which uses the average magnetic field strength of a patch, is not surprising.

\subsection{Inclination}                             
       
The kG magnetic fields found in the plage and network are typically vertical and we find no systematic difference between the two populations once canopy fields have been excluded. We do not expect any systematic differences between the two magnetic polarities and thus only the unsigned magnetic field inclination is analysed here. 

\begin{figure}
        \centering
        \includegraphics[width=8cm]{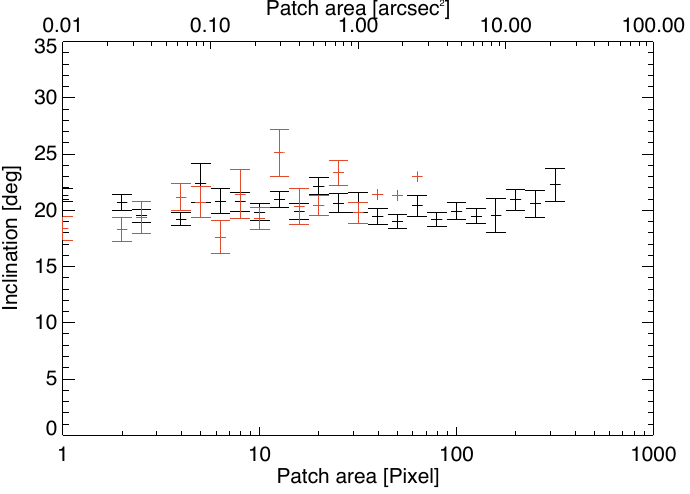} 
        \caption{Scatterplot of patch areas of kG feature and their mean inclination at $\log(\tau)=-0.8$. The black symbols refer to patches in the plage and the red symbols are from the network. The error bars refer to the error of the mean. The mean inclinations were calculated using ten logarithmic bins per decade of patch area.}
        \label{Incl}
        \end{figure}

Figure \ref{Incl} demonstrates that the average magnetic field inclination does not have a patch area dependence and there is no systematic difference between plage and network inclinations. The smallest patches in the network and plage areas display the largest deviation around the mean inclination of $20^{\circ}$ and some patches have a mean inclination of more than $40^{\circ}$. While the largest patches have average inclinations of $\sim20^{\circ}$, they nonetheless internally host pixels with inclinations larger than $40^{\circ}$, which are typically found near the edges of such a patch. Pixels with perfectly vertical magnetic fields are reliably found at the centre of pores and large patches in general. In addition, for a given patch area, we found no correlation between a patch's mean inclination and its overall magnetic field strength or continuum intensity.

We investigated the potential influence of noise on the retrieved inclinations by comparing data sets four and six in Table \ref{SPtable}. Data set four has a standard Stokes $Q$, $U$, and $V$ noise level of $1\times10^{-3}$ $I_c$, whereas data set six's noise level is $6\times10^{-4}$ $I_c$. However, the mean inclinations of kG pixels in data six are only $1^{\circ}-2^{\circ}$ more vertical across all $\log(\tau)$ layers compared to data set four and the standard deviations are identical, as well as their mean magnetic field strengths. 
         
\begin{figure}
        \centering
        \includegraphics[width=8cm]{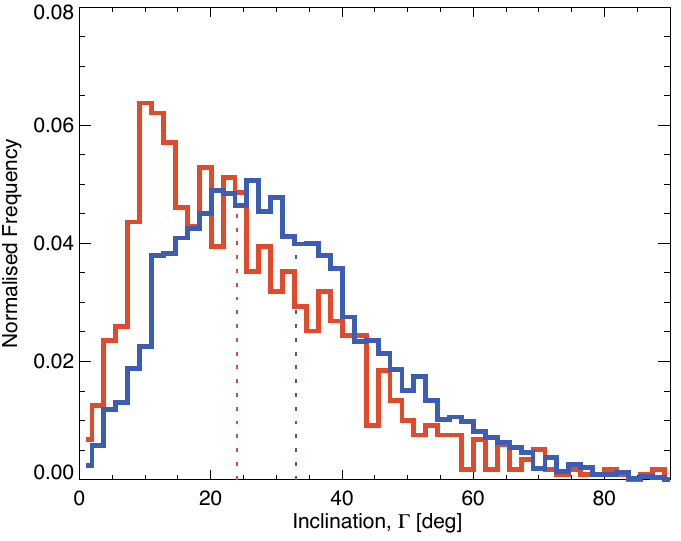}
        \caption{Histograms of inclination of canopy fields >500 G at $\log(\tau)=-2.0$ of kG features in areas of different magnetic flux density. The colour scheme is identical to Figure \ref{generalI}. The two dotted lines indicate the mean inclination of each distribution.}
        \label{CanIncl}
        \end{figure}

In plage areas, canopy fields cannot be readily associated with individual patches, as the canopies of neighbouring patches already merge at $\log(\tau)=-2$. Therefore, we are only able to make a general comparison between network and plage canopies. The red distribution in Figure \ref{CanIncl} displays the inclinations of canopy fields in the network and the blue for plage. The mean inclination of the network canopy is $24 \pm 4^{\circ}$ and $33 \pm 3 ^{\circ}$ for the plage canopy. We imposed an arbitrary cutoff field strength of 500 G at $\log(\tau)=-2$ for the distributions displayed in Figure \ref{CanIncl}. Lower cutoff fields strengths will gradually include more unrelated internetwork fields and will add more horizontal fields to both distributions. The colour scheme in Figure \ref{CanIncl} is identical to that of Figure \ref{generalI} and the canopy becomes gradually more inclined as the mean flux density in an area increases. The larger inclination of the canopy fields in areas with higher magnetic flux may be related to the areas' larger kG magnetic features. With increasing size the kG fields in the outer parts of the magnetic features tend to be more inclined, which also influence the canopy fields' inclination. Magnetic features which are located close to large magnetic structures such as sunspots typically possess canopies with a mean inclination larger than $50^{\circ}$. The plage canopy distribution in Figure \ref{CanIncl} does not change significantly if pixels located close sunspots are excluded from it indicating that the on average more horizontal plage canopy fields are not a result of their proximity to sunspots but intrinsic in nature. 
        
In addition, weak opposite polarity fields (< 400 G at $\log(\tau)=0$) that reside beneath the canopy of a kG patch were identified around network patches. They appear to be similar in nature to the weak opposite polarity fields beneath the canopies of patches in the plage described by \citet{buehler2015}.         
        
\subsection{Velocity}         

\begin{figure}
        \centering
        \includegraphics[width=8cm]{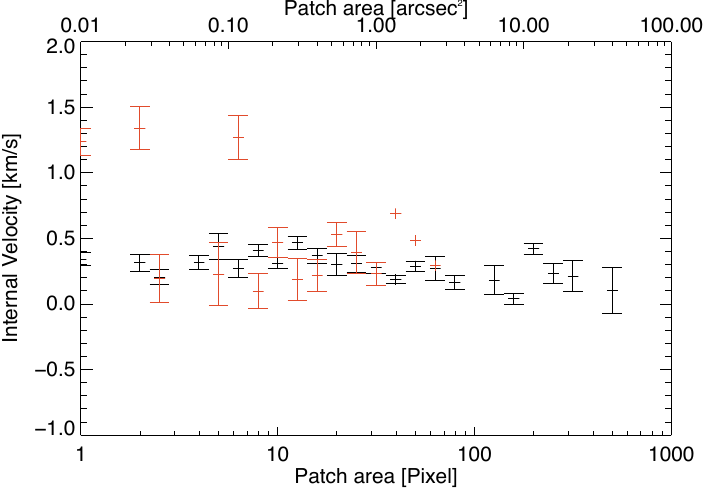} 
        \caption{Scatterplot of patch areas of kG features and mean internal flow speeds at $\log(\tau)=-0.8$. The black symbols refer to patches in the plage and the red symbols are from the network. The error bars indicate the error of the mean. The mean LOS velocities were calculated using ten logarithmic bins per decade of patch area.}
         \label{intvel}
         \end{figure}

The average LOS velocities within magnetic features in plage and network areas are usually no larger than 400 m/s across all patch areas as indicated by Figure \ref{intvel}. There appears to be no systematic difference between similarly sized patches found in the network and plage except for the smallest patches. In addition, the LOS velocities within the magnetic patches are not homogeneous. The average maximum upflows within a kG patch reach 1.5 km/s whereas the average maximum downflows attain 2 km/s. The increase of internal downflow speeds with decreasing patch size may partly have to do with the increasing difficulty of clearly separating internal flows from the surrounding ring of downflows in smaller features. It is therefore unclear to what extent the trend seen in Figure \ref{intvel} is real, or is affected by bias.

\begin{figure}
        \centering
        \includegraphics[width=8cm]{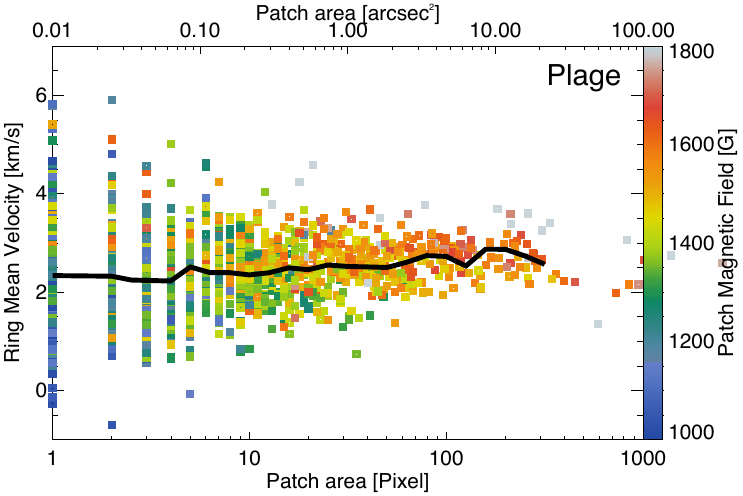} 
        \includegraphics[width=8cm]{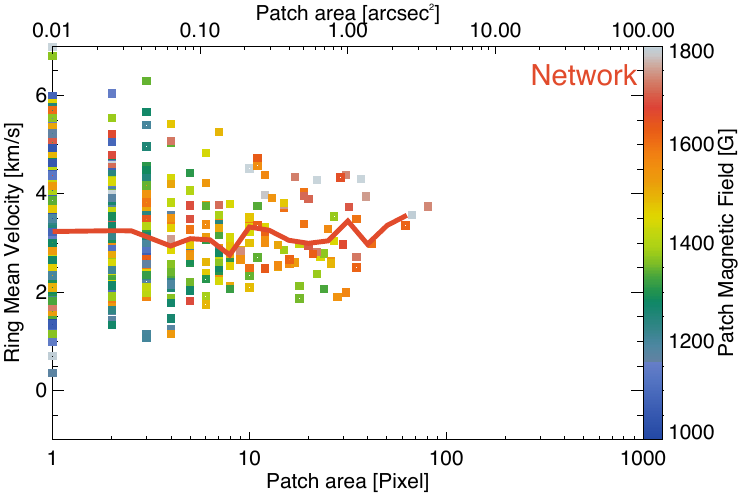}
        \caption{Scatterplot of patch areas of kG features and mean flow speeds at $\log(\tau)=0$ in a one pixel-wide ring surrounding them for features in the plage, top, and the network, bottom. The solid lines indicate mean flow speeds. The ring mean LOS velocities were calculated using ten logarithmic bins per decade of patch area.}
         \label{velavg}
         \end{figure}

The fast downflows surrounding the magnetic patches are faster than the downflow speeds measured in field-free intergranular lanes. We isolated these fast downflows by drawing a one pixel ring around each kG magnetic feature. The average velocity of each downflow ring around a magnetic feature at $\log(\tau)=0$ is plotted in Figure \ref{velavg}. The downflow rings host flows with speeds of 2.4 km/s around small <10 pixel plage patches but reach average speeds of 3.2 km/s around similarly sized features in the network, some 800 m/s faster.       
 
\begin{figure}
        \centering
        \includegraphics[width=8cm]{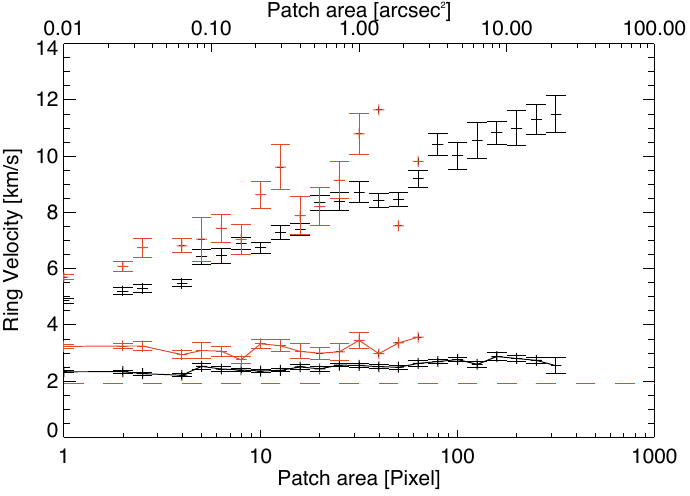} 
        \caption{Relation between kG feature size and flows surrounding them. The solid black line displays the mean flow speeds in a one-pixel ring surrounding kG features at $\log(\tau)=0$ in plage areas and the solid red line for network features. They are identical to the solid lines in Figure \ref{velavg}. The black plus symbols indicate the fastest single pixel flows in a ring in plage areas whilst red plus symbols show the same in network areas. The red dashed line indicates the mean flow in quiet Sun intergranular lanes. The errors bars denote the error in the mean. All the mean LOS velocities were calculated using ten logarithmic bins per decade of patch area.}
         \label{NTPvel}
         \end{figure}

The average downflow speed increases slightly with the size of the host magnetic feature, see solid lines in Figure \ref{velavg}, which are displayed again in Figure \ref{NTPvel} (black and red solid lines). The average maximum downflows also increase with patch area, only much more strongly, as indicated by the plusses in Figure \ref{NTPvel}. Each plus symbol is the average of the maximum downflow around all patches for a given patch area. Whilst the average maximum downflow around a bright point remains below 6 km/s at $\log(\tau)=0$, larger network features host faster downflows in individual pixels. The largest patches found in the plage, which typically contain pores, can host average maximum downflows, which are supersonic, exceeding 11 km/s.         

The downflow ring around a small patch such as a bright point contains on average $1\%$ of pixels with supersonic velocities, both in plage and in the network. The number of supersonic pixels in a downflow ring linearly increases with the logarithm of the patches size. A downflow ring of a 100 pixel patch hosts on average $4\%$ of pixels with supersonic velocities in the plage and $6\%$ in the network. Therefore, larger patches not only host the fastest supersonic downflows in their immediate surroundings, but also more of them. Quiet Sun intergranular lanes do not host supersonic downflows in any $\log(\tau)$ layer.       
         
\subsection{Microturbulence}

\begin{figure}
        \centering
        \includegraphics[width=8cm]{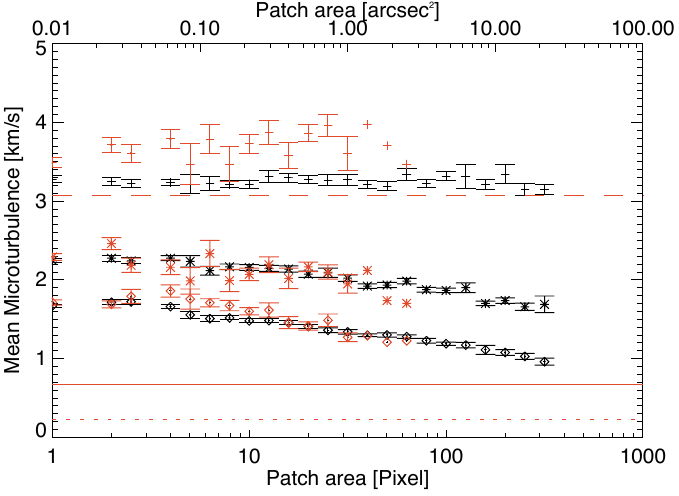} 
        \caption{Relation between kG feature size and its internal mean micro turbulent velocity. The plus, star and diamond symbols indicate the $\log(\tau)=0, -0.8$ and $-2$ layers respectively. The error bars denote the error in the mean. The black symbols refer to plage and the red to network areas. The dashed, solid, and dotted lines indicate the mean microturbulence in quiet Sun intergranular lanes at $\log(\tau)=0,-0.8$ and $-2$, respectively. All the mean microturbulent velocities were calculated using ten logarithmic bins per decade of patch area.}
         \label{NTPmicro}
         \end{figure}

The average microturbulence in magnetic patches in the network is generally higher than in the plage across all $\log(\tau)$ layers by some 300 - 500 m/s according to Table \ref{comp}. However, Figure \ref{NTPmicro} reveals that similarly sized features in the network and plage tend to have similar microturbulent velocities, except at $\log(\tau)=0$ where the network displays a larger microturbulence. The lower overall microturbulence in the plage appears to be caused by the largest patches, which are only present in plage. The largest patches in the plage often host pores of various sizes, which typically have lower microturbulent velocities in the upper $\log(\tau)$ layers. As a result, the smaller the magnetic feature, the larger the required microturbulence, except in the lowest layer, where the microturbulence is independent of patch area.

\begin{figure}
        \centering
        \includegraphics[width=8cm]{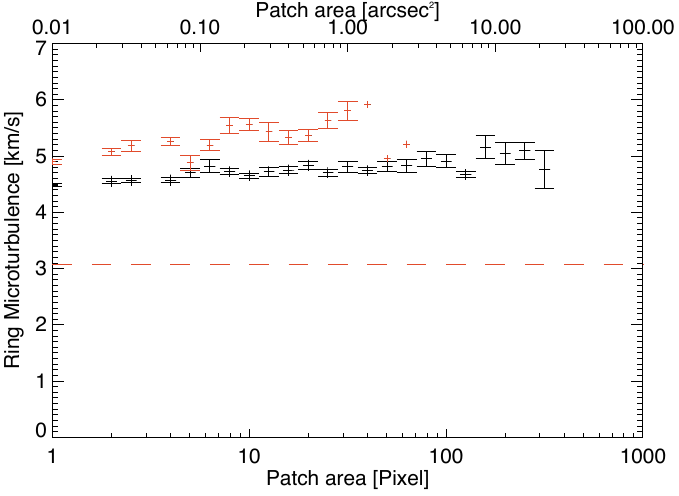} 
        \caption{Relation between kG feature size and mean micro turbulent velocity in a one pixel-wide ring surrounding them for features in the plage, black, and the network, red at $\log(\tau)=0$. The error bars denote the error in the mean. The dashed line indicates the mean microturbulence in quiet Sun intergranular lanes at $\log(\tau)=0.$ The mean microturbulent velocities were calculated using ten logarithmic bins per decade of patch area.}
         \label{NTPmicroring}
         \end{figure}

The highest microturbulent velocities are typically found within the ring of downflows that surrounds each magnetic feature. Figure \ref{NTPmicroring} indicates that at $\log(\tau) =0$ the micro turbulent velocities in the rings can reach average speeds of 4.5 km/s for bright points and even 5 km/s for the largest patches in the plage. The average maximum microturbulent speeds in a single pixel in a ring ranges from 7 km/s up to 10 km/s for the largest features. The patch area dependence of the microturbulent velocities in Figure \ref{NTPmicroring} is qualitatively similar to the LOS velocities in the rings in Figure \ref{NTPvel}. Furthermore, the microturbulent velocities in the rings are higher than, both, within the kG patches, as well as in the quiet Sun indicated by the horizontal line in Fig \ref{NTPmicroring}.  
 
\section{Discussion}

In the previous section we examined several properties of kG magnetic features in plage and network areas. In this section we will interpret our results and compare them to the literature.

The magnetic fields in plage and network areas modify the continuum intensity in a complex way. Weak hG fields are found in intergranular lanes or on granules \citep{khomenko2003} and typically have continuum intensities close to the mean quiet Sun intensity \citep{schnerr2011}. Stronger kG magnetic fields by comparison can host continuum intensities in excess of granular intensities in the form of bright points, or far below intergranular intensities in the form of pores. In plage areas, where magnetic features of all sizes are present, plotting $B$ vs $I$ (Figure \ref{BT}) reveals that the continuum intensity peaks at an intermediate magnetic field strength before rapidly dropping off for larger magnetic field strengths, which is consistent with previous studies \citep{lawrence1993,kobel2011}. Unlike previous studies we find our peak intensity at 1.1 kG, which is higher than previously reported. Network areas do not display a drop off in continuum intensity for stronger magnetic fields, which qualitatively compares well with simulated 'red' continuum images by \citet{danilovic2013} and \citet{criscuoli2013} and the simulated G-band observations of \citet{riethmueller2017} as well as the $525$ nm continuum observations by \citet{kahil2017}. However, both the average and individual intensities of magnetic features in the network are higher than in previous observations \citep[e.g.][]{berger2007,kobel2011,kahil2017} using comparable wavelengths and spatial resolution due to our 2D inversion approach.

Both network and plage areas host kG magnetic features with an average magnetic field strength of $1.5$ kG at $\log(\tau)=-0.8$, which compares favourably with earlier investigations \citep{zayer1989,keller1990,lin1995,martinez1997}. The magnetic features can be approximated by bright thin fluxtube models, but individual magnetic patches often host multiple kG cores \citep{berger2004,requerey2015}. Larger magnetic features, such as pores, host stronger magnetic fields compared with smaller features such as bright points, in accord with, for example, \citet{zayer1990}. Unlike previous investigations we find that kG features in the network have higher magnetic field strengths than similarly sized features in the plage (see Figure \ref{Field}). This results appears to contradict the conclusion of \citet{stenflo1985}, who found that areas of high magnetic flux host stronger kG fields. Although the line ratio technique employed by \citet{stenflo1985} is able to differentiate kG features from quiet Sun areas within a resolution element, they were not able to separate the canopy fields from the central cores of their kG features, as we have done in this investigation. We are able to reproduce the result of \citet{stenflo1985} when we include canopy fields in the calculation of the average magnetic field in plage and network areas. Plage areas have on average at least 100 G stronger fields compared to network areas when canopy fields are included in the calculation, in agreement with \citet{stenflo1985}.

The analysis of 3D MHD simulations by \citet{roehrbein2011} and \citet{danilovic2013} indicate that the continuum intensity of bright points steadily increases with increasing magnetic field strength. Our results confirm this finding and demonstrate that bright points in the network with high magnetic field strength produce a stronger continuum intensity enhancement than bright points with lower magnetic field strengths, in line with earlier 2D MHD simulations \citep{schuessler1986,knoelker1988}. Our results indicate that bright points in network host on average higher magnetic fields ($\sim$150 G, see Figure \ref{Field}) and higher continuum intensities ($\sim$5\%, see Figure \ref{BTpatch}) than  equally large plage counterparts. The observations by \citet{romano2012} using the G-band images with a smaller pixel scale of $0\farcs085$ support our results whereas \citet{ji2016}, using intensity images obtained from TiO observations at 706 nm, reported the opposite behaviour where the intensity contrast appears to correlate with the mean magnetic field strength. However, they only selected bright points in both plage and network samples and used SDO/HMI magnetograms to obtain the mean magnetic field strength. The comparatively low pixel scale of $0\farcs505$ of the HMI magnetograms are likely to produce higher magnetic field strengths for plage bright points, since plage regions generally feature clusters of bright points adjacent to pores resulting in a kG filling factor of closer to unity in plage areas than in the quiet Sun. \citet{Liu2018} find by comparing two $20'' \times 20''$ boxes of differing magnetic flux that the bright point intensity at 706 nm is on average independent of magnetic flux. It is likely that the two boxes used are too small a sample to accurately ascertain the dependence of bright point intensity on magnetic flux as the scatter in Figure \ref{generalI} suggests.

Areas of high average magnetic flux, such as plage, feature an overall decreased continuum intensity at disc centre compared to the quiet Sun, which stems not only from the presence of pores \citep{peck2019} but also from the modified and less efficient convection in such areas \citep[see][and Figs. \ref{generalI} and \ref{contrast}]{ishikawa2007,kobel2012} in accord with MHD simulations \citep{voegler2005,criscuoli2013}. In network areas the enhanced brightness from kG features is able to outshine the reduced continuum intensity from its field-free regions compared to a quiet Sun with no kG features. Therefore, even at disc centre kG features in the network cause an overall excess brightness of $0.1\%$ compared to a more field free part of the quiet Sun, which supports the evidence presented by \citet{yeo2013} and \citet{criscuoli2017} using full-disc HMI data. Furthermore, the stray-light corrected HMI images employed by \citet{criscuoli2017} support the evidence presented by \citet{kobel2011} and our investigation that magnetic field concentrations are brighter in network than in plage areas.

In addition, the modified convection in plage areas may also change the efficiency of the convective collapse mechanism \citep{proctor1982,venkatakrishnan1986}, which may serve as an explanation for the reduced field strength and continuum intensity of bright points in the plage compared with features of the same size in the network.

The kG features in both, plage and network areas host predominantly vertical magnetic fields with average inclinations of $20^{\circ}$ and mode inclinations between $10^{\circ}-20^{\circ}$ in agreement with \citet[][]{bernasconi1995,buehler2015} and Figure \ref{Incl}. Additionally, the average inclination of kG features is patch area independent, but smaller patches in particular display a larger dispersion around the mean, which may be an indication of granular buffeting \citep{steiner1996}. In addition, we report that the small, weak opposite polarity patches residing beneath the canopies of kG features described by \citet{buehler2015} are also present around kG features in the network.

However, the inclination of magnetic fields in the canopies of kG features are on average $9^{\circ}$ more horizontal in plage areas than in network areas (see Figure \ref{CanIncl}). The more horizontal canopy magnetic fields in the plage stem from the presence of large magnetic features which expand faster than their smaller counterparts. Such large features are absent in the network. Plage features situated in the immediate vicinity of larger magnetic flux concentrations such as sunspots display fields that are even more inclined, both in their cores and their canopies \citep{buehler2015}. 

The kG magnetic features within plage and network areas host weak average downflows of $400$ m/s or less across all $\log(\tau)$ layers (see Figure \ref{intvel}), which agrees well with previous studies \citep{stenflo1984,solanki1986,martinez1997,buehler2015}. We consider the obtained internal downflows as an upper limit, as there is probably some residual cross-talk from the ring of rapid downflows surrounding the magnetic features into the internal pixels (see next paragraph).  The $\pm1.5$ km/s internal up and downflows in kG features, excluding the fast dowflows located at the edges of the features, are possibly transient in nature and may be an indication of flux tube waves \citep{solanki1992wave}. 

Magnetic features are additionally surrounded by fast downflows, which typically exceed $2$ km/s at $\log(\tau)=0$  (see Figure \ref{velavg}) and match previous observations \citep{rimmele2004,langangen2007,buehler2015} and both, theoretical and empirical models \citep{deinzer1984,grossmann1988,solanki1989,buente1993,voegler2005}. We can expand upon these earlier findings and report that magnetic features in the network are surrounded by flows that are on average $800$ m/s faster than flows found around similarly sized magnetic features in plage. The faster flows in the network are likely caused by the more efficient convection in network areas \citep{narayan2010,kobel2012}. Fast downflows around pores have previously been reported \citep{keil1999,stangl2005,cho2010} and we can generalise this phenomenon to all kG magnetic features. Furthermore, we find that the magnitude of the surrounding downflows is correlated with the logarithm of the patch area and magnetic field strength of the kG patch (see Figure \ref{NTPvel}). The rapid downflows known to be surrounding pores are in line with this relationship.

Magnetic features in the network and plage host internal microturbulent velocities \citep{holweger1978}. Investigations by \citep{solanki1986} and \citet{zayer1990} reported on average larger turbulent velocities in network  than in plage areas with both being larger than the turbulent velocity in the quiet Sun (they used a combination of micro- and macro-turbulence to reproduce the line profiles). We confirm these results in the present study (see Figure \ref{NTPmicro}). However, we also show that similarly sized magnetic features require similar internal microturbulent velocities regardless of their location. The lower average microturbulence in plage areas results from a patch area dependence whereby larger features require smaller microturbulent velocities than smaller ones. The small pores, which are found in plage areas, are nearly devoid of microturbulence at $\log(\tau)=-0.8$ and $-2$ and thereby reduce the overall average microturbulent velocities required in plage. The high microtrbulence required by the smallest kG features may be an indication that they are still not completely resolved by our observations. Interestingly, the large microturbulence at $\tau=1$ is nearly independent of patch area. It may be a sign of unresolved magnetoconvection, which is typically restricted to the lower $\log(\tau)$ layers. The fact that the microturbulent velocity in the magnetic features is higher than in the quiet Sun and drops off more slowly is also unexpected in terms of (magneto-)convection, which is expected to be weaker in kG magnetic elements than in the quiet Sun. The results may be amenable to another interpretation: A part of the microturbulent velocity may be due to short-wavelength, high frequency waves within the magnetic features. Waves with wavelengths of the order of the width of the contribution function of a spectral line mainly lead to line broadening, very similarly to a micro-turbulence.
The addition of more spectral lines from different atoms in the inversion may aid in the resolution of this issue and further constrain the inversion parameters as has been demonstrated in the photosphere by \citet{riethmuller2018} and in the chromosphere by \citet{dasilva2018}.

\section{Conclusion}

The properties of kG magnetic features found in plage and network areas are compared using 2D SPINOR inversions of six SP disc centre scans aboard Hinode SOT. The seeing-free high resolution SP data, combined with the power of the 2D SPINOR inversions, allow us to extend such comparisons beyond what has been possible in the past.

The average kG magnetic feature in both network and plage areas expands like a thin flux tube hosting 1.5 kG magnetic fields, an inclination from the vertical of $10^{\circ}-20^{\circ}$, slow internal downflows typically below 400 m/s, and fast downflows exceeding 2 km/s surround the magnetic features. Magnetic features in plage areas are on average some 200 K cooler and host smaller microturbulent velocities compared to their network counterparts. All the average properties of these features agree with previously reported values. The large microturbulent velocities compared with the quiet Sun, in particular in the higher layers, may be an indicator of unresolved internal motions, such as caused by short-wavelength waves.

A more detailled inspection of our inversion results revealed that magnetic features in the network have on average 150 G stronger magnetic fields and $5\%$ higher continuum contrasts than similarly sized structures in plage. In addition, magnetic features in the network are surrounded by downflows, which are on average $800$ m/s faster than in the plage at $\log(\tau)=0$. The maximum magnitude of the downflows is feature size dependent, consequently the largest magnetic features in the plage produce the fastest maximum downflows in their immediate surroundings. The maximum downflows can exceed 11 km/s around the largest magnetic patches, which is faster than any previously reported photospheric flow outside of a sunspot. The magnetic canopy in plage areas is on average $9^{\circ}$ more horizontal than in the network, which may partially explain the different chromospheric structuring in network and plage areas. It may explain the preponderance of spicules in active regions. Small, weak magnetic fields residing beneath the canopies of kG features can be found in both the network and plage areas.

\begin{acknowledgements}
Hinode is a Japanese mission developed and launched by ISAS/JAXA, with NAOJ as domestic partner and NASA and STFC (UK) as international partners. It is operated by these agencies in co-operation with ESA and NSC (Norway).
This project has received funding from the European Research Council (ERC) under the European Union's Horizon 2020 research and innovation programme (grant agreement No. 695075) and has been supported by the BK21 plus program through the National Research Foundation (NRF) funded by the Ministry of Education of Korea.
\end{acknowledgements}

\bibliographystyle{aa}
\bibliography{/Users/davidbuehler/Documents/Latex/TheBib_copy}{} 

\begin{appendix}

\section{Overview of the data}

        \begin{figure*}
        \centering
        \includegraphics[width=13cm]{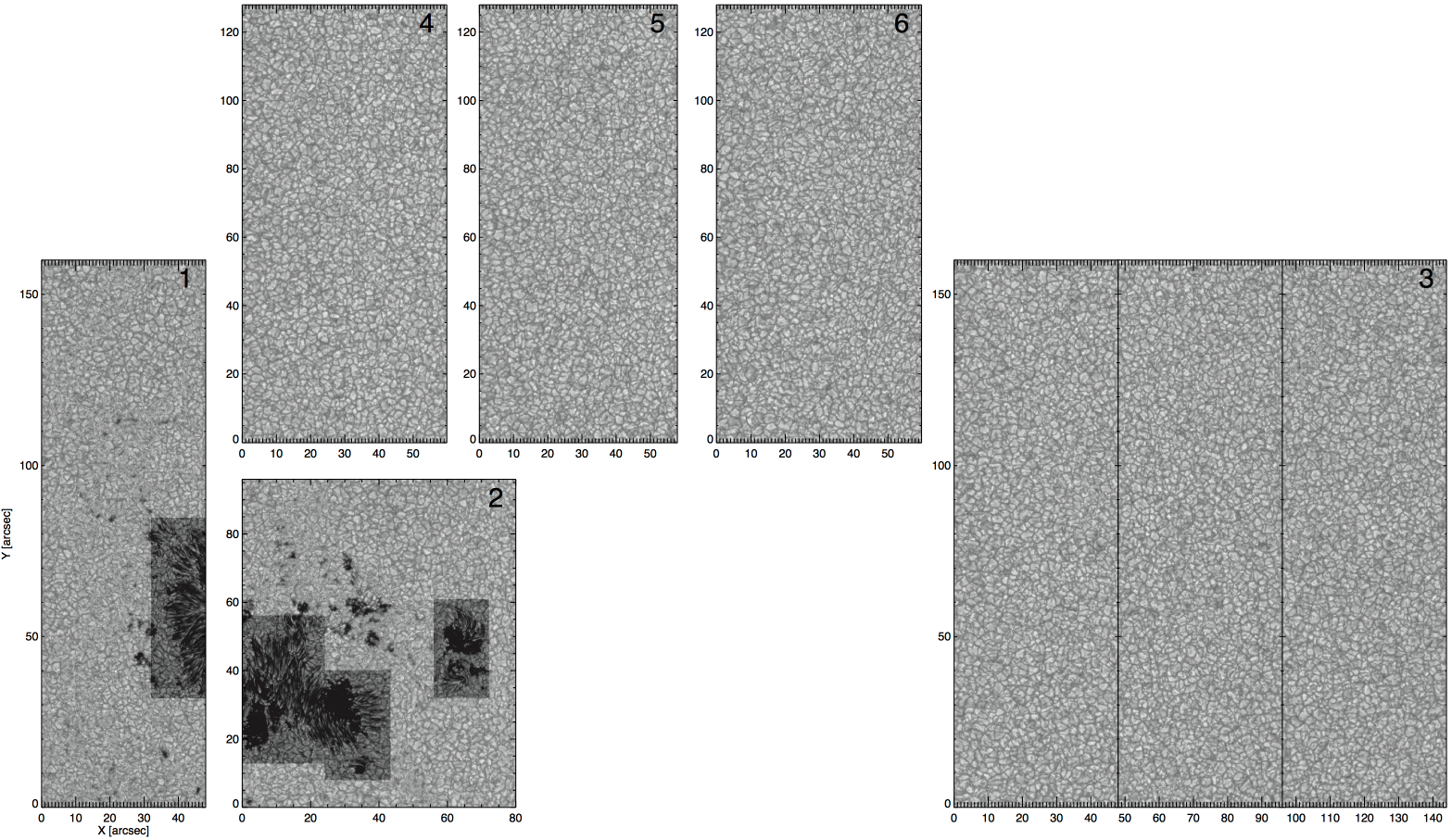}
        \caption{Continuum intensity images of the data sets listed in Tab \ref{SPtable}. The shaded areas have been excluded from the analysis.}
        \label{overviewNTPcont}
        \end{figure*}
        
        \begin{figure*}
        \centering
        \includegraphics[width=13cm]{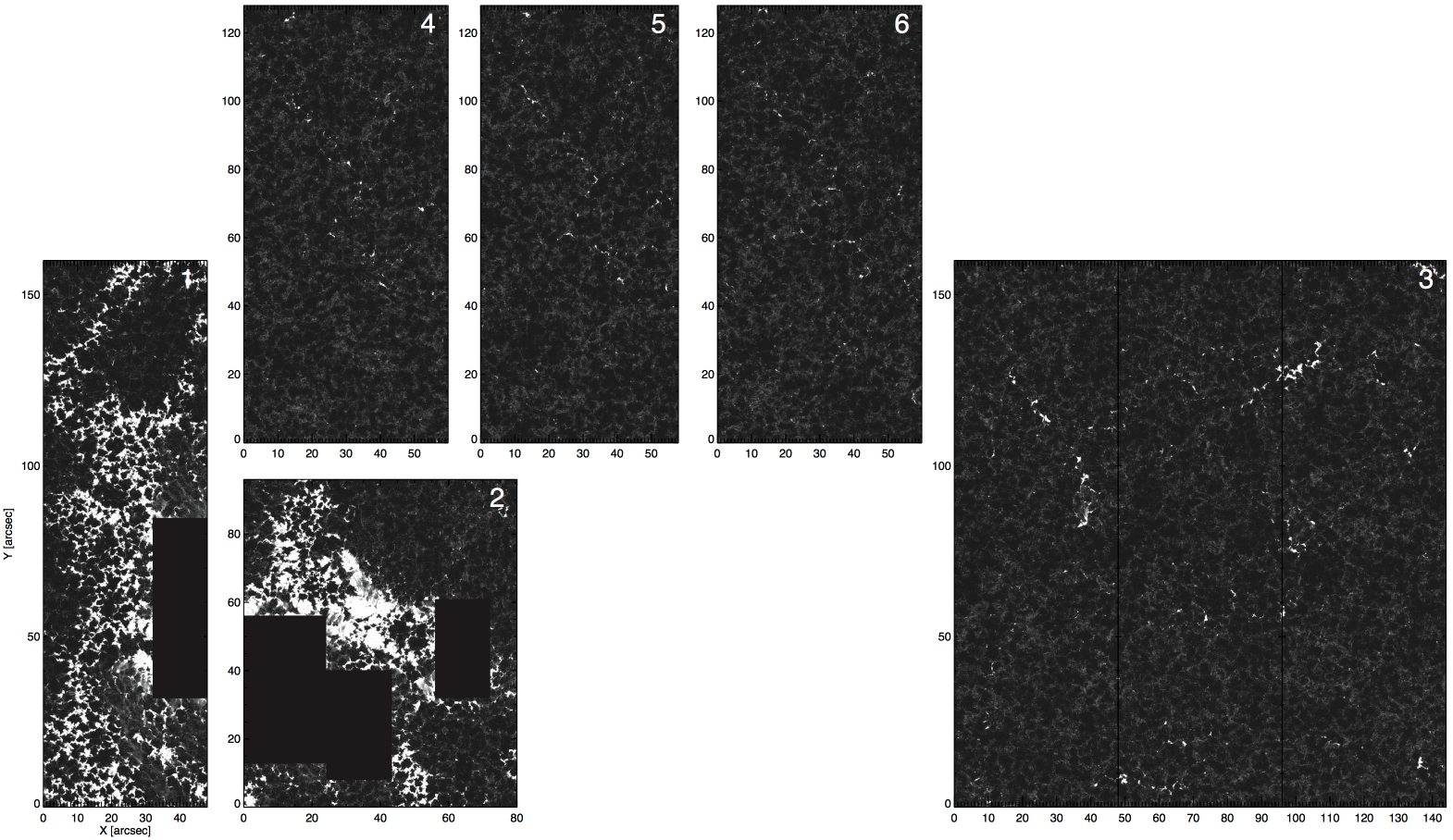}
        \caption{Images of the magnetic field strength at $\log{\tau}=-0.8$ of the data sets listed in Tab \ref{SPtable}. Areas that have been excluded from the analysis are blackened out.}
        \label{overviewNTPmag}
        \end{figure*}
        
\end{appendix}

\end{document}